\documentclass[superscriptaddress,showpacs,twocolumn]{revtex4}
\usepackage{amssymb}
\usepackage[tbtags]{amsmath}
\usepackage{graphicx}
\usepackage{epsfig,graphicx,times}

\begin{document}

\title{Probing tiny motions of nanomechanical resonators: classical or quantum mechanical?}
\author{L.F. Wei}
\affiliation{Frontier Research System, The Institute of Physical
and Chemical Research (RIKEN), Wako-shi, Saitama, 351-0198, Japan}
\affiliation{IQOQI, Department of Physics, Shanghai Jiaotong
University, Shanghai 200030, China}
\author{Yu-xi Liu}
\affiliation{Frontier Research System, The Institute of Physical
and Chemical Research (RIKEN), Wako-shi, Saitama, 351-0198, Japan}
\author{C.P. Sun}
\affiliation{Frontier Research System, The Institute of Physical
and Chemical Research (RIKEN), Wako-shi, Saitama, 351-0198, Japan}
\affiliation{Institute of Theoretical Physics, The Chinese Academy
of Sciences, Beijing, 100080, China}
\author{Franco Nori}
\affiliation{Frontier Research System, The Institute of Physical
and Chemical Research (RIKEN), Wako-shi, Saitama, 351-0198, Japan}
\affiliation{Physics Department, MCTP, CSCS, The University of
Michigan, Ann Arbor, Michigan 48109-1040}
\date{\today }

\begin{abstract}
We propose a spectroscopic approach to probe tiny vibrations of a
nanomechanical resonator (NAMR), which may reveal classical or
quantum behavior depending on the decoherence-inducing
environment. Our proposal is based on the detection of the
voltage-fluctuation spectrum in a superconducting transmission
line resonator (TLR), which is {\it indirectly} coupled to the
NAMR via a controllable Josephson qubit acting as a quantum
transducer. The classical (quantum mechanical) vibrations of the
NAMR induce symmetric (asymmetric) Stark shifts of the qubit
levels, which can be measured by the voltage fluctuations in the
TLR. Thus, the motion of the NAMR, including if it is quantum
mechanical or not, could be probed by detecting the
voltage-fluctuation spectrum of the TLR.
\end{abstract}


\maketitle

\textit{Introduction.---\,} Since the beginning of quantum theory,
many researchers have tried to monitor macroscopic quantum effects
with mechanical resonators (see, e.g.,~\cite{BO96}). This relates
to the debate on the quantum-classical mechanics boundary for
macroscopic objects and the mechanisms of quantum
decoherence~\cite{CL}. Besides superconductivity and Bose-Einstein
condensates, quantum oscillations of nanomechanical resonators
(NAMRs) could also provide an attractive platform for
experimentally testing quantum phenomena at macroscopic scales.
Furthermore, reaching the quantum limit of mechanical motions
could open new avenues of technology~\cite{Cho03}, in, e.g., high
precision measurement, quantum computation, and even gravitational
wave detection.

A mechanical resonator may reveal either quantum or classical
behavior, depending on the decoherence-inducing
environment~\cite{CL}. Phenomenologically (see, e.g., Ref. [4]),
if the energy ($h\nu$) of the vibration (with frequency $\nu$)
quanta is larger than the thermal energy $k_B T$, then the
mechanical oscillation could be regarded as quantum mechanical.
NAMRs with low thermal occupation number have recently been
experimentally studied~\cite{Gaidarzhy05,LaHaye04}. These
nanodevices, containing $10^{10}$---$10^{12}$ atoms, work at very
low temperatures (in the mK-range) and sufficiently high
frequencies (GHz-range), approaching the quantum limit. A
formidable challenge (see, e.g.,~\cite{Gaidarzhy05,LaHaye04}) in
this field is how to sensitively detect the quivering of the
detected nanodevice, and {\it quantitatively} verify whether it is
quantum mechanical or not. Indeed, it is difficult to {\it
directly} detect~\cite{LaHaye04,Bruder} the tiny displacements of
a NAMR, vibrating at GHz frequencies, using the available
displacement-detection techniques. Also, the usual
position-measurement method is ultimately limited by the
always-present ``zero-point motion" fluctuations in the quantum
regime~\cite{BO96}.

Here, we propose a promising {\it indirect\/} method to detect the
mechanical oscillation of a NAMR approaching its quantum limit.
Instead of attempting to further improve the sensitivity of the
usual force/displacement detection~\cite{LaHaye04} or to redesign
the tested nanostructure~\cite{Gaidarzhy05}, our proposal is based
on the detection of the voltage-fluctuation spectrum in a
superconducting transmission line resonator (TLR). A controllable
Josephson qubit, acting as a quantum electro-mechanical
transducer~\cite{sun05}, is used to couple the NAMR to the TLR.
Our approach is conceptually similar to that in quantum optics for
verifying the field quantization in a cavity~\cite{Raimond96}, and
provides a {\it quantitative test\/} to distinguish the two types
of mechanical motions: either {\it quantum or classical}.
Namely, compared to the spectrum of the TLR without a NAMR, the
{\it classical} motion of the NAMR only {\it symmetrically}
increases the vacuum Rabi splitting, while the {\it quantum}
motion of the NAMR {\it further shifts} the positions of the peaks
to the right. Physically, this difference originates from the
commutativity of the classical variables $\alpha$ and $\alpha^*$,
for classical oscillators, as opposed to the noncommutativity of
the corresponding bosonic operators $\hat{b}$ and
$\hat{b}^\dagger$ for quantum oscillators. Thus, for large
detuning, the classical (quantum) NAMR symmetrically
(asymmetrically) shifts the qubit levels. The symmetric shifts
enlarge the vacuum Rabi splitting symmetrically, and the {\it
additional} displacement of the excited level in the asymmetric
Stark shifts, induced by the {\it quantum} NAMR, further shifts
the peaks to the right.

\textit{Model.---}\,We consider a simple circuit quantum
electrodynamics (CQED) system~\cite{Blais04,Wallraff04}
schematically sketched in Fig.~1. A Josephson
qubit~\cite{Shnirman97}, formed by two Cooper-pair boxes connected
via two identical Josephson junctions (with capacitance $c_J$ and
Josephson energy $\varepsilon_J$), is capacitively coupled to a
TLR (of total capacitance $C_t$, length $L$), via a capacitance
$C_0$, and an electrostatically-modulated NAMR (of mass $m$ and
frequency $\omega_R$), via a capacitance $C_x=C_d(1+x/d)^{-1}$.
The oscillating NAMR (driven, e.g., by an external force pulse)
modulates the gap (with displacement $x$ around the equilibrium
distance $d$), and thus the coupling capacitance $C_x$ between the
NAMR plate and the bottom Cooper-pair box. Here, $C_d$ is the gate
capacitance between the non-oscillating NAMR plate (corresponding
to $x=0$) and the bottom Cooper-pair box, which is biased by the
gate-voltage $V_g$ via the gate capacitance $C_g$. We assume
$C_J=2c_J \ll C_0=C_d=C$ to safely neglect the {\it direct}
interaction between the NAMR and the TLR; their {\it indirect}
connection is realized by simultaneously coupling to the common
qubit, acting as a switchable quantum transducer. The total excess
Cooper-pair number $n_t$ in the two boxes (the bottom ``b" and
upper ``u" ones) is $n_t=n_b+n_u=1$; and
$|\downarrow\rangle=|n_b=1,n_u=0\rangle$ and
$|\uparrow\rangle=|n_b=0,n_u=1\rangle$ are the two typical charge
states. Near the degenerate point (i.e., $V_x+V_g\approx0$), this
device~\cite{Shnirman97} forms a good two-level artifical ``atom",
described by the pseudo-spin operators $\sigma_z=|e\rangle\langle
e|-|g\rangle\langle g|$, $\sigma_+=|e\rangle\langle g|$, and
$\sigma_-=|g\rangle\langle e|$, with
$|g\rangle=\cos(\alpha/2)|\uparrow\rangle+\sin(\alpha/2)|\downarrow\rangle$
and
$|e\rangle=-\sin(\alpha/2)|\uparrow\rangle+\cos(\alpha/2)|\downarrow\rangle$,
and $\tan\alpha=E_J/\omega_0$. The ``atomic" eigenfrequency
$\omega_0=(E_{C}^2+E_{J}^2)^{1/2}$ could be controlled by the
applied gate voltages $V_g,\, V_x$, and the biasing external flux
$\Phi_e$. In fact, $E_{C}=eC(V_g+V_x)/(2C_{J}+C)$ and
$E_J=2\varepsilon_J\cos(\pi\Phi_e/\Phi_0)$, with $\Phi_0=h/2e$.

The Hamiltonian of our CQED system can be written as
\begin{equation}
H=H_S+\nu\hat{a}^\dagger\hat{a}+\lambda(\sigma_+\hat{a}+\sigma_-\hat{a}^\dagger)
+H_{\rm TLR-bath}+H_{\rm q-bath},
\end{equation}
with $\hbar=1$.
Depending on the different motions of the NAMR, the first term in
Eq.~(1) takes the different forms:
(i) $H_S=\omega_0\sigma_z/2=H_N$ for the no-oscillation case
``$N$"\,---\,when the NAMR plate does {\it not} oscillate; (ii)
$H_S=H_N+\zeta[\sigma_+\exp(-i\omega_R t)+\sigma_-\exp(i\omega_R
t)]=H_C$ for the classical case ``$C$"\,---\,the NAMR plate
oscillates classically with frequency $\omega_R$;\, and (iii)
$H_S=H_N+\omega_R\hat{b}^\dagger\hat{b}+
\zeta(\sigma_+\hat{b}+\sigma_-\hat{b}^\dagger)=H_Q$ for the
quantum case ``$Q$"\,---\,the NAMR plate oscillates
quantum-mechanically with frequency $\omega_R$, respectively. All
higher-order terms of $x/d$ have been neglected~\cite{sun04}, as
the quivering $x$ of the NAMR is sufficiently small (compared to
$d$), e.g., $ x/d\sim 10^{-6}$.
The second- and third terms in Eq.~(1) describe a selected bare
mode with frequency $\nu$ in the TLR and its coupling
($\propto\lambda$) to the qubit. The coupling strengths $\lambda$
and $\zeta$, listed above, are
$\lambda=-\sqrt{\nu/C_t}\,eC\sin\alpha/(2C_{J}+C)]$ and
$\zeta=\sqrt{1/(2m\omega_R)}\,eCV_x\sin\alpha/[2d(2C_{J}+C)]$,
respectively. Under the usual rotating-wave approximation, we have
also neglected the rapidly-oscillating terms
$\sigma_{-}\exp(-i\omega_R t),\,\sigma_{+}\exp(i\omega_R t)$ (in
the couplings of the qubit to the classical NAMR),\,
$\sigma_{-}\hat{b},\,\sigma_{+}\hat{b}^\dagger$ (in the couplings
of the qubit to the quantum-mechanical NAMR), and
$\sigma_{+}\hat{a}^\dagger,\,\sigma_{-}\hat{a}$ (in the
interaction between the qubit and the TLR).
Dissipation in the NAMR determines~\cite{CL} the vibrational modes
of the NAMR: classical or quantum mechanical, and thus the form of
$H_S$. While dissipation in the selected TLR mode and the
Josephson qubit {\it directly} influences the voltage-fluctuations
in the TLR. Here, we describe these two dissipations via the last
two terms of Eq.~(1): $H_{\rm {\rm TLR
}-bath}=\sum_{j}(\omega_j\,\hat{c}_j^\dagger\,\hat{c}_j+
u_j\,\hat{c}_j\,\hat{a}^\dagger+u_j^*\,\hat{c}_j^\dagger\,\hat{a})$
and $H_{\rm q-bath}=\sum_{k}(\omega_k\,\hat{d}_k^\dagger\,
\hat{d}_k+
v_k\,\hat{d}_k\,\sigma_++v_k^*\,\hat{d}_k^\dagger\,\sigma_-)$,
with $\{\hat{c}_j,\,\hat{c}_j^\dagger,\,j=1,2,3,...\}$ and
$\{\hat{d}_k,\,\hat{d}_k^\dagger,\,k=1,2,3,...\}$ being the
corresponding bosonic operators of two independent reservoirs:
$c$-bath and $d$-bath, respectively. Also, $u_j$\,(or $v_k$) is
the coupling between the selected TLR mode (or qubit) and the
$j$th (or $k$th) mode of the $c$- (or $d$-) bath.

A central motivation of the present work is to detect the motion
of the NAMR by measuring the correlation spectrum
%\begin{widetext}
\begin{eqnarray}
S_V(\omega)&=&\frac{1}{2\pi}\int_{-\infty}^{+\infty}d\tau\,e^{i\omega\tau}\,\langle\hat{V}(y,t)\hat{V}(y,t+\tau)
\rangle_{t\rightarrow\infty}\\
&\propto&\int_0^{+\infty}\hspace{-0.3cm}dt_1\int_0^{+\infty}\hspace{-0.3cm}dt_2\,\exp[i\omega(t_2-t_1)]\,\langle\hat{a}^\dagger(t_1)\hat{a}(t_2)\rangle
\nonumber
\end{eqnarray}
%\end{widetext}
of the voltage $V(y,t)$ at site $y$ (e.g., $V(L,t)=V_{\rm out}(t)$
in Fig.~1) in the TLR. The second line in Eq.~(2) comes from the
fact that the voltage $V(y,t)$, contributed by the selected mode
of frequency $\nu$ along the TLR, is quantized~\cite{Blais04};
$\hat{V}(y,t)\propto[\hat{a}^\dagger \exp(-i\nu
t)+\hat{a}\exp(i\nu t)]$.
We estimate that the voltage-signal in the TLR is sufficiently
strong, and can be measured by using a standard rf network
analyzer~\cite{Gaidarzhy05}. Indeed, the voltage amplitude, even
for the fundamental-mode vacuum fluctuation of the typical
TLR~\cite{Wallraff04}, is up to $V_{\rm rms}=\sqrt{\nu/C_t}\sim
2\,\mu$V, corresponding to an electric field $E_{\rm rms}\sim
0.2$\,V/m, which is much larger than that in the usual optical
$3$D atom-QED system~\cite{Raimond96}.

\begin{figure}[tbp]
\vspace{-0.5cm}
\includegraphics[width=12cm]{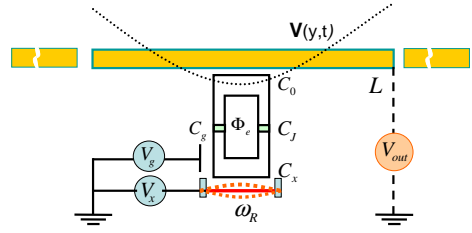}
\vspace{-5.6cm} \caption{(Color online). Schematic diagram of a
nanomechanical resonator (NAMR) (dashed red lines, with vibrating
frequency $\omega_R$) indirectly coupled to a superconducting
transmission line resonator (TLR), shown in yellow, of length $L$
(with voltage distribution $V(y,t)$ shown by the black dotted-line
on top) via a Josephson qubit with small junction capacitances.
The upper (lower) Cooper-pair box of the qubit capacitively
couples to the TLR (NAMR), via a capacitance $C_0$\,\,($C_x$). The
voltage-fluctuation spectrum of $V_{\rm out}$, at the right end of
the TLR, reads-out motional information of the NAMR.}
\end{figure}

\textit{Spectra of the TLR.---\,} If the bare TLR (without
coupling to the qubit) is excited at a selected mode of frequency
$\nu$, the measured voltage-spectrum should have a Lorentzian
shape~\cite{Wallraff04,scully}: $S_0(\omega)\propto
1/[(\omega-\nu)^2+(\gamma/2)^2]$, centered at $\nu$ and with a
width at half-height of $\gamma=\nu/Q_{\nu}$, corresponding to the
finite quality factor $Q_{\nu}$ of that mode due to its
dissipation.

First, we consider the voltage-fluctuation spectrum $S_N(\omega)$
of the TLR coupled to the qubit, in the absence of NAMR
oscillations. In this case $H_S=H_N$, and the the system is
initially prepared in the state
$|\Psi(0)\rangle=|e\,0_a0_c0_d\rangle$, i.e., the qubit is in its
excited state $|e\rangle$, the field mode and baths are in the
vacuum states~\cite{note}: $|0_a0_c0_d\rangle=|0_a\rangle\otimes
|0_c\rangle\otimes|0_d\rangle$, with
$|0_c\rangle=\prod_{j=1}^{\infty}|0_j\rangle,\,
|0_d\rangle=\prod_{k=1}^{\infty}|0_k\rangle$, respectively. The
wavefunction, $|\Psi\rangle=|{\rm
qubit,photon,TLR-bath,q-bath}\rangle$, of the system at arbitrary
time $t$ takes the form~\cite{scully,Law95}
\begin{eqnarray}
|\Psi(t)\rangle&=&c_1(t)|g\,1_a0_c0_d\rangle+c_2(t)|e\,0_a0_c0_d\rangle\\
&+&\sum_{j=1}^{\infty}C_j(t)|g\,0_a\{1_j\}0_d\rangle
+\sum_{k=1}^{\infty}D_k(t)|g\,0_a0_c\{1_k\}\rangle,\nonumber
\end{eqnarray}
with $|\{1_j\}\rangle=|1_j\rangle\otimes\prod_{j'\neq
j}|0_{j'}\rangle$ and
$|\{1_k\}\rangle=|1_k\rangle\otimes\prod_{k'\neq
k}|0_{k'}\rangle$.  Thus, the measured voltage-spectrum is
determined by the time-dependence of $c_1(t)$, i.e.,
\begin{equation}
\langle\hat{a}^\dagger(t_1)\hat{a}(t_2)\rangle=c_1^*(t_1)c_1(t_2).
\end{equation}
Without loss of generality and for simplicity, we assume that the
qubit is adjusted to resonance with one of the eigenmodes of the
TLR~\cite{Wallraff04}, e.g., $\omega_0=\nu=2\pi\times 6$\,GHz.
Then, under the usual Weisskopf-Wigner
approximation~\cite{scully}, the desirable voltage-fluctuation
spectrum can be calculated as
\begin{eqnarray}
S_N(\omega)\propto \left(\frac{\lambda}{\Delta_N}\right)^2
\left|A_+^{-1}-A_-^{-1}\right|^2,
\end{eqnarray}
with $A_{\pm}=-(\gamma_c+\gamma_d)/4+
i[\omega-(\nu\mp\Delta_N)/2]$, and
$\Delta_N=\sqrt{4\lambda^2+\gamma_c\gamma_d-(\gamma_c+\gamma_d)^2/4}$.
This $S_N(\omega)$ is a spectrum with a two-peak structure; each
peak has a width at half height of $(\gamma_c+\gamma_d)/2$, and
the distance between peaks is the vacuum Rabi splitting
$\Delta_N$. Above, $\gamma_c$ and $\gamma_d$ are the damping rates
of the qubit excited state and the selected TLR mode,
respectively.

Second, after preparing the present CQED system (biased by a
non-zero gate-voltage $V_x$) in the initial state
$|\Psi(0)\rangle$, we drive the NAMR to oscillate mechanically by
a force pulse and then measure the voltage-fluctuation spectrum of
the TLR. Usually, the interaction between the NAMR and the qubit
works in the large-detuning regime~\cite{Armour02}:
$\eta=\zeta/\delta\ll 1$, i.e., $\zeta\ll
\delta=\omega_0-\omega_R$. In this limit, the NAMR oscillation
does {\it not} change the qubit-state populations, and only
results in {\it Stark shifts on the qubit levels}.
Indeed, neglecting higher-order small quantities $O(\eta^2)$, the
Hamiltonians $H_C$ and $H_Q$ can be effectively
approximated~\cite{wei05} to
$H_S^{(C)}=(\omega_0/2+\zeta^2/\delta)\sigma_z$ and
$H_S^{(Q)}=\omega_0\sigma_z/2+\zeta^2(n_c\sigma_z+|e\rangle\langle
e|)/\delta$, respectively. Here, $n_c$ is the quantum occupation
number of the quantum-mechanical NAMR. $H_S^{(C)}$ implies that,
if the NAMR oscillation is classical, two energy levels of the
qubit experience {\it symmetric} (i.e., equivalent) Stark shifts:
upward for $|e\rangle$ and downwards for $|g\rangle$. Thus, the
case $C$ is really similar to the non-oscillator case N discussed
above, except that now the modified qubit is not in resonance with
the selected TLR mode.
However, if the NAMR oscillation is {\it quantum-mechanical},
i.e., for the case $Q$, the Stark {\it shifts} shown in
$H_S^{(Q)}$ for the two levels of the qubit are {\it no} longer
equivalent (i.e., {\it asymmetric}). Namely, the energy increase
of $|e\rangle$ (Stark shift) is different from the energy decrease
in $|g\rangle$. Thus, the tiny motions of the NAMR could be
probed, via $S_V(\omega)$, by detecting the above NAMR-induced
Stark shifts of the qubit levels.

Since the NAMR (now oscillating in the large-detuning regime) does
not induce any quantum transition in the circuit, the wavefunction
at $t>0$ of the system with NAMR still takes the form in Eq.~(3).
However, the voltage-fluctuation spectrum of the TLR will change
to
%\begin{widetext}
\begin{eqnarray}
S_C(\omega)\propto
\left(\frac{\lambda}{\Delta_C}\right)^2\left|B_+^{-1}-B_-^{-1}\right|^2,
\end{eqnarray}
%\end{widetext}
with $B_\pm=-(\gamma_c+\gamma_d)/4\pm \xi_C/2+i[\omega-(\nu\mp
\chi_C)/2],\,\xi_C=\Delta_C\sin(\theta_C/2),\,\chi_C=\Delta_C\cos(\theta_C/2)$,
for the classical case $C$; and
\begin{eqnarray}
S_Q(\omega)\propto
\left(\frac{\lambda}{\Delta_Q}\right)^2\left|C_+^{-1}-C_-^{-1}\right|^2,
\end{eqnarray}
%\end{widetext}
with $C_\pm=-(\gamma_c+\gamma_d)/4\pm
\xi_Q/2+i[\omega-(\nu+\zeta^2/\delta\mp
\chi_Q)/2],\,\xi_Q=\Delta_Q\sin(\theta_Q/2),\,\chi_Q=\Delta_Q\cos(\theta_Q/2)$,
for the quantum case $Q$, respectively.
Above,
$\Delta_l=\left([4\lambda^2+\varrho_l^2+\gamma_c\gamma_d-(\gamma_c+\gamma_d)^2/4]^2
+\rho_l^2(\gamma_c-\gamma_d)^2]\right)^{1/4}$,\,
$\theta_l=\arctan[\rho_l(\gamma_c-\gamma_d)/(4\lambda^2+\varrho_l^2+\gamma_c\gamma_d-(\gamma_c+\gamma_d)^2/4)]$,
($l=C,\,Q$), and $\varrho_C=2\zeta^2/\delta$,
$\varrho_Q=(2n_c+1)\zeta^2/\delta$.
\begin{figure}[tbp]
\vspace{-1.3cm}
\includegraphics[width=4.2cm, height=5cm]{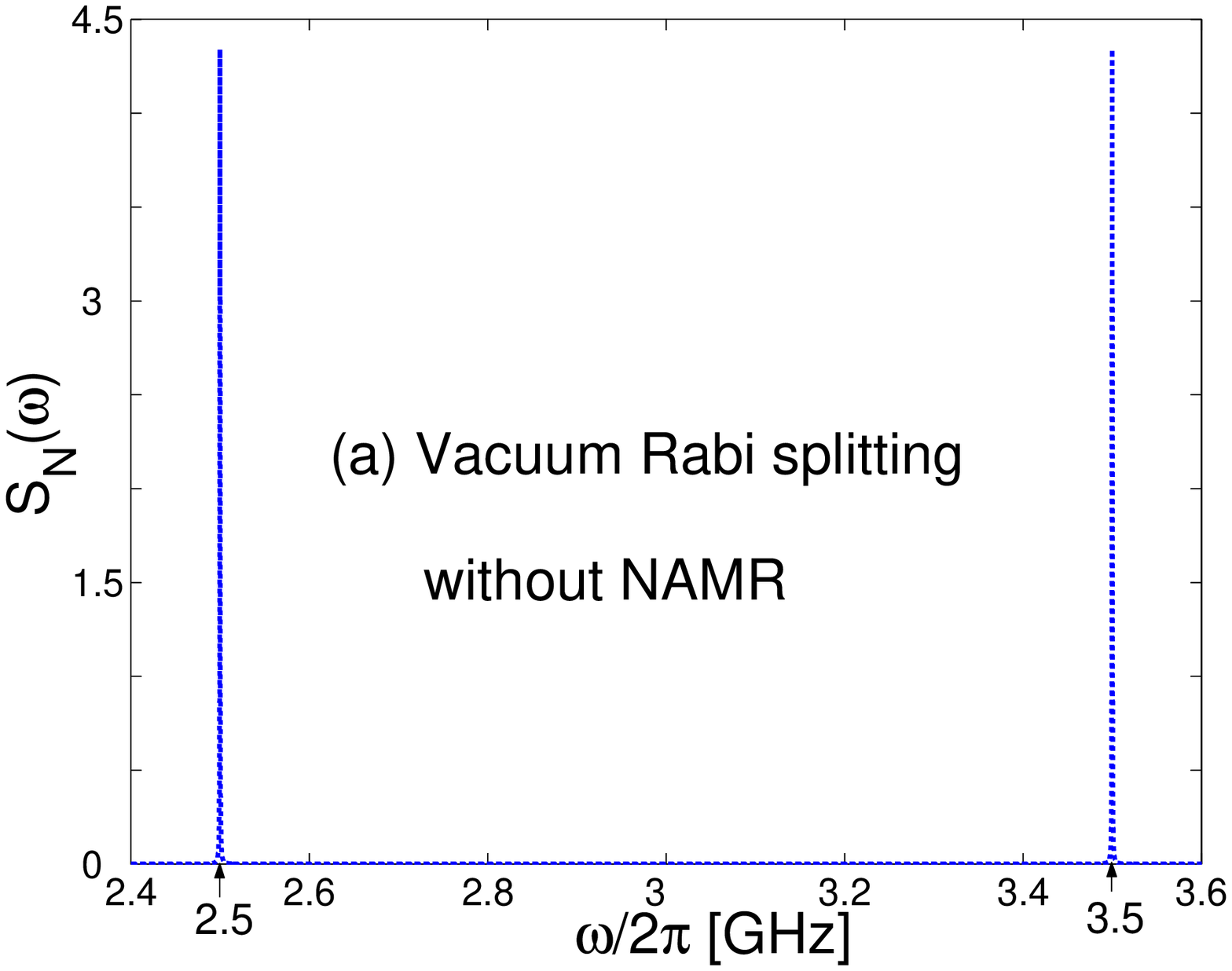}
\includegraphics[width=4.2cm, height=5cm]{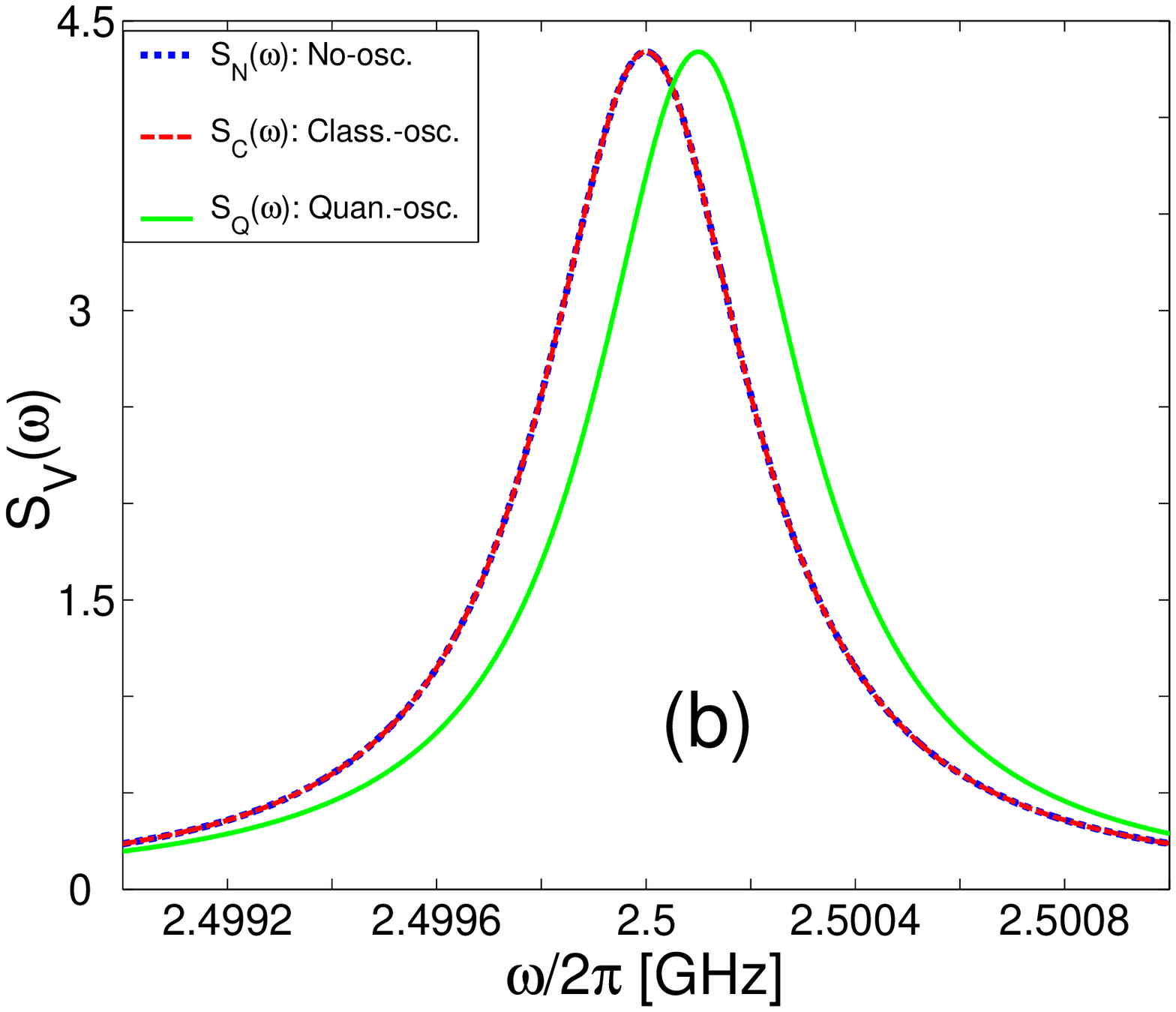}
\vspace{-1.2cm} \caption{(Color online). Voltage-fluctuation
spectra $S_V(\omega)$ of the TLR: (a) vacuum Rabi splitting in the
absence of the NAMR vibration, (b) The modifications of the left
peak in $S_N(\omega)$ due to the vibrations of the NAMR in the
weak coupling case: $\zeta^2/\delta=200$\,kHz. The red dashed-line
$S_C(\omega)$ (on top of $S_N(\omega)$) corresponds to the
classical NAMR. The green solid-line $S_Q(\omega)$ (distinguished
from $S_N(\omega)$ by a shift to the right) corresponds to the
quantum mechanical NAMR with occupation number $n_c=1$.}
\end{figure}
%
%\vspace{2cm}

In the present strong-coupling CQED system, $2\lambda\gg \gamma_c,
\gamma_d$ and $\theta_l\sim 0$, thus, when the NAMR does not
oscillate, the two peaks of the measured spectrum $S_N(\omega)$
are approximately at $\omega=\nu/2\pm\Delta_N/2$ with the vacuum
Rabi splitting $\Delta_N\approx 2\lambda$.
The classically oscillating NAMR shifts the positions of the two
peaks in $S_N(\omega)$ to $\omega\approx (\nu/2\pm \Delta_C/2)$
and enlarges the vacuum Rabi splitting from $\Delta_N$ to
$\Delta_C$, with an additional splitting $\Delta_C-\Delta_N\approx
\varrho^2_C/(4\lambda)=\zeta^4/(\lambda\delta^2)$. While, if the
oscillation of the NAMR is quantum mechanical, not only the vacuum
Rabi splitting is enlarged (from $\Delta_N$ to $\Delta_Q$) by an
increment $\Delta_Q-\Delta_N\approx
\varrho^2_Q/(4\lambda)=(n_c+1/2)^2\zeta^4/(\lambda\delta^2)$, but
also the positions of the two peaks are shifted to the right by
$\Delta\omega=\zeta^2/(2\delta)$ to $\omega\approx
\nu/2\pm\Delta_Q/2+\Delta\omega$.
%
%The reason why these peaks shift, is that, besides the symmetric
%Stark shifts on the qubit levels (induced by the classical
%motion), the quantum NAMR induces an additional displacement (of
%$2\Delta\omega$, coming from the last term of $H_S^{(Q)}$) of the
%excited level, which shifts the two peaks to the right by an
%additional amount $\Delta\omega$.

For typical parameters (e.g.,~\cite{Wallraff04,LaHaye04,Armour02}
$Q_{\nu}=10^4$ for $\nu=\omega_0=2\pi\times 6$ GHz,
$\omega_R=2\pi\times 1$\,GHz, $C_J/C\sim 0.1$, $\zeta=2\pi\times
30$\,MHz, and $\lambda\sim 2\pi\times 500$
MHz,\,$\gamma_d=0.6\,\gamma_c$), Fig.~2(a) shows the vacuum Rabi
splitting of the TLR spectrum $S_N(\omega)$ in the absence of the
NAMR. Figure 2(b) shows how the NAMR mechanical oscillations
modify the voltage-fluctuation spectrum in the TLR. There, we only
show how the left peak of $S_N(\omega)$ is shifted in the presence
of the NAMR coupled to the qubit. The shift of the right peak can
be analyzed similarly. Obviously, the vibration of the NAMR
modifies the level-structure of the Josephson qubit, and thus
changes the voltage-fluctuation spectral distribution of the TLR:
from $S_N(\omega)$ to either $S_C(\omega)$ or $S_Q(\omega)$,
depending on the motional features of the NAMR oscillation:
classical or quantum mechanical. For the case when there is weak
coupling between the possible existing NAMR oscillation and the
qubit (e.g., $x/d\sim 1.0\times 10^{-6}$, yielding
$\zeta^2/\delta\sim 200$\,kHz in Fig.~2(b)), the effect of
increasing the vacuum Rabi splitting is very weak:
$\Delta_B-\Delta_N\approx\Delta_C-\Delta_N\sim 80$\,\,Hz, which
may not be easily detectable.
However, even in such a weak coupling, the effect of shifting the
peak of $S_N(\omega)$ to the right, due to the quantum mechanical
NAMR oscillations, should be detectable:
$\Delta\omega=\zeta^2/(2\delta)\sim 2\pi\times 100$\,KHz.

Given the experimental parameters $\omega_0(=\nu),\omega_R$, and
$\lambda$, a small decrease of $d$ may yield a large increase in
the coupling $\zeta$, and thus the effects discussed above may be
much stronger: as $\Delta_l-\Delta_N\propto\zeta^4$ and
$\Delta\omega\propto\zeta^2$. Figure 3 shows the modification of
$S_N(\omega)$ due to the qubit driven by a strongly-coupled NAMR
with $x/d\sim 7.1\times 10^{-6}$, yielding $\zeta^2/\delta\sim
10$\,MHZ, and thus $\Delta\omega\sim 5$\,MHz. In this case, both
the classical and quantum mechanical NAMR can be detected.
Compared to the left peak of $S_N(\omega)$, the left peak of
$S_C(\omega)$ has been left shifted with a quantity $\chi_C/2\sim
100$\,KHz, just due to the increment $\chi_C$ of the vacuum Rabi
splitting. While, if the quantum mechanical NAMR is coupled to the
qubit, then the left peak of the spectrum $S_N(\omega)$ will be
shifted to the left with $\chi_Q/2$ due to the increased vacuum
Rabi splitting $\chi_Q$, and shifted to the right with
$\Delta\omega=\zeta^2/2\delta$. The net result is that this peak
will be shifted to the right by $\Delta\omega-\chi_Q/2\approx
2\pi\times 4.8$\,MHz, and thus the left peak of $S_Q(\omega)$
would be now centered at $\nu/2+\Delta\omega-\chi_C/2\approx
2\pi\times2504.8$\,MHz. This shift could be easily detected.

\begin{figure}[tbp]
\vspace{-1.4cm}
\includegraphics[width=8cm, height=6cm]{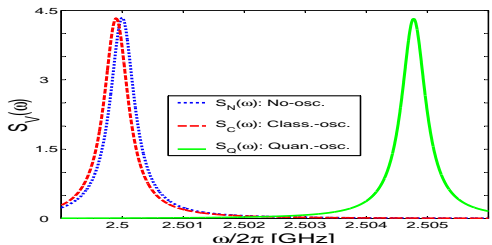}
\vspace{-1.5cm}\caption{(Color online). Shifts of the left peak of
$S_N(\omega)$ (see Fig.~2(a)) when the qubit-NAMR coupling becomes
stronger than in the case shown in Fig.~2(b). Here,
$\zeta^2/\delta=10$\,MHz. In this case, the left peak of
$S_N(\omega)$ undergoes a small (large) shift to the left (right)
by the classical (quantum-mechanical with $n_c=1$) vibrations of
the NAMR.}
\end{figure}

\textit{Conclusion and Discussions.---}The tiny oscillations of a
NAMR should reveal either quantum or classical behavior. We have
proposed an effective approach to test this by {\it indirectly}
probing it.
This is because different types of motion of the NAMR would induce
different Stark shifts on the qubit levels, and thus modify
differently the spectrum of the TLR.
Our proposal is experimentally realizable. It is possible, at
least in principle, to fabricate the sufficiently small Josephson
capacitance $C_J$ for realizing the indirect coupling between the
NAMR and the TLR, via a commonly connected Josephson qubit. Also,
the mechanical motions of the NAMR in current
experiments~\cite{LaHaye04} are approaching the quantum limit, and
satisfy the large-detuning condition required in the present
proposal. In fact, $\omega_R\lesssim 1$\,GHz, $\omega_0=\nu\sim
6$\,GHz in current experiments~\cite{Wallraff04,LaHaye04}, and we
estimate $\zeta\sim 30$\,MHz (for $C_J/C\sim 0.1$ and $V_x\sim
0.1$\,V). This implies that $\eta=\zeta/\delta\,\sim 6\times
10^{-3}\ll 1$.

Dissipation exists in the NAMR~\cite{Cleland02,note2}, i.e, its
quality factor $Q_R$ is finite. However, even for the weak
NAMR-qubit coupling discussed above (e.g., $\zeta\sim 2\pi\times
30$\,MHz), and a relative low quality factor~\cite{Cleland02},
e.g., $Q_R=10^3$, the decay $\gamma_R=\nu/Q_R$ of the NAMR is
still very small: $\gamma_R/\zeta\sim 1/30$. Thus, our proposed
test, based on the observation of shifts in the peaks of the
voltage spectrum, is not strongly affected by dissipation.

This work is partially supported by the NSA, LPS, ARO, ARDA, AFOSR
contract No. F49620-02-1-0334, and the NSF grant No. EIA-0130383.


\begin{thebibliography}{99}
\bibitem{note}At experimentally accessible temperatures, e.g., $T\sim 30$ mK~\cite{Wallraff04},
the baths can be initialized to their ground states: the mean
number of thermal photons (in the microwave region of wavelength
$0.1$ --- $15$ cm) are negligible, i.e., $\langle
n_{th}\rangle\approx 10^{-208}$--- $0.043 \ll 1$.
\end{thebibliography}
\end{document}